\documentclass[a4paper,11pt]{article}
\usepackage[utf8]{inputenc}
\usepackage{color}
\usepackage{cite}
\usepackage{hyperref}
\usepackage{amsmath}
\usepackage{graphicx}

\usepackage[top=0.4in, bottom=0.7in, left=0.60in, right=0.5in]{geometry}

\begin{document}
\title{{\Large Radiative $E1$ transitions between $^3P_1$ and $^3S_1$ quarkonium states}}
\author{Vaishali Guleria$^1$, Eshete Gebrehana$^2$, Shashank Bhatnagar$^1$}
\maketitle \small{1. Department of Physics, University Institute of Sciences, Chandigarh University, Mohali-140413, India\\}
\maketitle \small{2. Department of Physics, Woldia University, Ethiopia\\}

\begin{abstract}
\normalsize{In this work we study the E1 decay processes, $^3P_1$ $\rightarrow$ $^3S_1\gamma$, and $^3S_1$ $\rightarrow$ $^3P_1\gamma$ in the framework of Bethe-Salpeter equation and calculate their decay widths. We have used algebraic forms of Salpeter wave functions obtained through analytic solutions of mass spectral equations for ground and excited states of $^3S_1$, and $^3P_1$ equal mass quarkonia in approximate harmonic oscillator basis to do analytic calculations of their decay widths. These decay widths have been compared with data and other models.}
\end{abstract}

\section{Introduction} Radiative transitions offer valuable insights into the dynamics and properties of quarkonia. Understanding the electromagnetic transitions in quarkonia is crucial to the field of theoretical and experimental particle physics, shedding light on the fundamental forces that govern the behavior of these exotic particles. Further, among quarkonium decays, radiative decays are particularly valuable as tests of various models, since the photon is directly observed and the nature of the electromagnetic transition is well understood. $E1$ transitions between $Q\bar{Q}$ states are the simplest radiative transitions of the form $J^{++}\rightarrow 1^{--}\gamma$ (with $J=0,1,2$). And since the E1 transitions are characterized by $|\Delta L|=1$, thus in these transitions there is a change in parity between the initial and final hadronic states. Electric dipole transitions are much stronger than magnetic dipole transitions, and involve transitions between excited states.

The present work involves radiative E1 transitions involving $^3S_1$ and $^3P_1$ charmonium states. We wish to mention that though a lot of investigation has been carried out on S-wave vector mesons such as, $0^{-+}$, and $1^{--}$, however comparatively lesser investigation has been carried out on P-wave mesons such as $1^{++},1^{+-}$ and $0^{++}$. Now, regarding the $1^{++}$ mesons, they were first seen in pp collisions by R704 Collaboration. However, not many decays of these mesons are experimentally observed as can be checked from PDG tables \cite{workman22,zyla20,olive14}. In 2013, the Belle Collaboration\cite{bhardwaj13} reported measurements of $B\rightarrow \chi_{c1}\gamma K$ decays, and $B\rightarrow \chi_{c2}\gamma K$ decays, where the $\chi_{c1}$ and $\chi_{c2}$ decay to $J/\Psi \gamma$. These results were obtained from a data sample of $776 \times 106$ $B\bar{B}$ events collected with the Belle detector at the KEKB asymmetric-energy $e^- e^+$ collider operating at the $Y(4S)$ resonance.

Further, an indirect way of producing P-wave states is through $e^- e^+$ annihilation, which first produces $^3 S_1$ charmonium states such as $J/\Psi$, and $\Psi(2S)$, followed by their radiative M1 and E1 decays which produce charmonium states, $^1S_0$ (such as $\eta_c$), and $^3P_1$ states (such as $\chi_{c1}$). We list here some of the indirect ways of producing $1^{++}$ state:  In \cite{coan06}, from $e^- e^+$ collision data  acquired with the CLEO detector at CESR, they observed the non-$D\bar{D}$ decay $\Psi(3770)\rightarrow \gamma\chi_{c1}$ with a statistical significance of $6.6\sigma$, using the two-photon cascades to $J/\Psi$ and $J/\Psi\rightarrow l + l$. We further wish to mention that in 2015, the BESIII collaboration\cite{ablikim15} independently confirmed the existence of the charmonium-like state $X(3823)$ in the $\chi_{c1}\gamma$ system with a statistical significance of $6.2\sigma$. This observation was made in the $e^-e^+$ collision process, specifically $e^-e^+\rightarrow\pi^+ \pi^-\chi_{c1}\gamma$ \cite{ablikim15}. The measured mass of $X(3823)$ in this study is $3823.7\pm1.8\pm0.7$ MeV\cite{ablikim15}, which is consistent with the mass measurement previously reported by the Belle Collaboration, thus providing further validation for the existence of the $X(3823)$ particle. The characteristics of $X(3823)$ align with $1^3D_2$ state, which is now officially recognised as $\Psi_2(3823)$\cite{workman22}. The primary decay modes of $\Psi_2(3823)$ are expected to involve radiative and hadronic transitions into other charmonium states. According to phenomenological studies, it is anticipated that the partial widths for $\Gamma_{\psi_2\rightarrow \chi_{c1}\gamma}$ are in the range of approximately 200 to 300 keV, while $\Gamma_{\psi_2\rightarrow \chi_{c2}\gamma}$ is estimated to be around 60 keV. Additionally, $\Gamma_{\Psi_2\rightarrow \ J/{\psi}\pi\pi}$ is predicted to have a partial width of approximately 160 keV.

However, it's worth noting that experimental results from BESIII have provided upper limits, indicating that the ratios of $\Gamma_{\Psi_2(3823)\rightarrow J/{\Psi}(\pi\pi)}$ to $\Gamma_{\Psi_2(3823)\rightarrow \chi_{c1}(\gamma)}$ are less than 0.06 and $\Gamma_{\Psi_2(3823)\rightarrow  J/{\Psi}(\pi^0\pi^0)}$ to $\Gamma_{\Psi_2(3823)\rightarrow \chi_{c1}\gamma}$ are less than 0.11. These experimental limits are in stark contrast to the theoretical predictions. This suggests that the decay channel $\psi_2\rightarrow \chi_{c1}\gamma$ may be the dominant mode for $\psi_2$ decay. In experimental observations, LHCb has established an upper limit of $\Gamma_{\psi_2}<5.2MeV$, while a more recent measurement by BESIII has refined this constraint to $\Gamma_{\psi_2}< 2.9$ MeV with $90\%$ confidence.

Also, recently, \cite{ablikim18} searched for processes, $e^- e^+\rightarrow \phi \chi_{c0, c1, c2}$ using a data sample collected with the BESIII detector operating at the BEPCII storage ring at a centre-of-mass energy, $\sqrt{s}= 4.6$ GeV. They further observed the processes, $e^- e^+\rightarrow \phi \chi_{c1}$, and $e^- e^+\rightarrow \phi \chi_{c2}$, with a statistical significance of more than $10\sigma$.

Now, we have earlier studied the  mass spectrum, and the leptonic decays of ground and excited states of heavy-light axial vector mesons such as $1^{++}$ and $1^{+-}$ in \cite{vaishali21a}, along with the radiative E1 decay processes\cite{vaishali21} involving $1^{+-}$ mesons, such as, $1^{+-}\rightarrow 0^{-+}\gamma$, and $0^{-+}\rightarrow 1^{+-}\gamma$.  Thus in the present work, we were interested in calculating the radiative decay processes involving $1^{++}$ mesons such as, $1^{++}\rightarrow 1^{--}\gamma$, and $1^{--}\rightarrow 1^{++} \gamma$. Here we wish to mention that radiative decays $^3P_1$ $\leftrightarrow$ $^3S_1$ receive contributions from both the $E1$ and $M2$ transitions, where $M2$ transitions have obvious corrections to the radiative transitions of $\chi_{c1}\rightarrow J/\Psi \gamma$, by interfering with the E1 transitions. The corrections from the $M2$ transition might be $(10 - 20)\%$ \cite{ylshi16,deng16} of the total partial decay widths.

There have been a number of theoretical studies of these E1 transitions based on a range of different models\cite{ebert03,godfrey15,segovia16,he21,he19}. Radiative transitions between $1^{--}$ and $1^{++}$ which proceed through the emission of a photon, are characterized by $|\Delta L|=1$, Thus in these transitions there is a change in parity between the initial and final hadronic states, though the total parity remains conserved.
We determine the radiative decay widths of heavy-light quarkonia for the processes mentioned above using a $4\times 4$ Bethe-Salpeter equation (BSE). This BSE approach is fully relativistic, taking into account the relativistic effects of quark spins and providing a consistent treatment of the internal motion of constituent quarks within the hadron due to its covariant structure. We have wave functions that fulfill the three-dimensional (3D) Bethe-Salpeter equation (BSE). This equation is derived from the four-dimensional (4D) BSE by employing the covariant instantaneous ansatz (CIA),which is a Lorentz-invariant extension of the instantaneous approximation. Consequently, our wave functions incorporate relativistic effects.

This paper is organized as follows: In Sec. II, we give the general formulation of the process ${H\rightarrow H'+\gamma}$ in the framework of a $4\times 4$ Bethe-Salpeter equation under the covariant instantaneous ansatz. In Sec. III, we calculate
the single photon decay widths for the process $A^+\rightarrow V\gamma$ . In
Sec. IV, we deal with the process $V\rightarrow A^+\gamma$. In Sec. V, we give numerical results and discussions.

\section{Radiative decays of quarkonia through $V\rightarrow A^+\gamma$}
Radiative M1 and E1 decays proceed through the process, $H\rightarrow H'\gamma$, where $H$ and $H'$ are initial and final quarkonia, that can be considered as two Lorentz frames. At leading order, two triangle quark-loop diagram shown in Fig.1  contribute to this process. The second diagram is obtained from the first by reversing the direction of internal fermion lines. We calculate this process in the $4\times 4$ representation of BSE.   We had earlier calculated some of the M1 and E1 decays of quarkonium in \cite{bhatnagar20,vaishali21}. Thus we will retain the same notation as in \cite{bhatnagar20,vaishali21}.

\begin{figure}[h!]
 \centering
 \includegraphics[width=12cm,height=6cm]{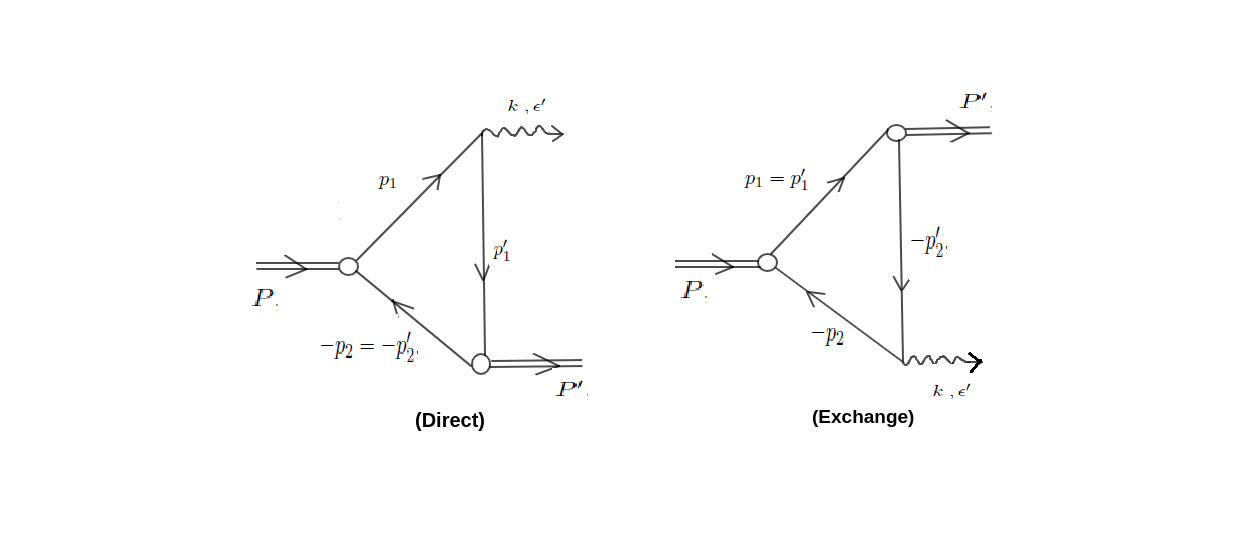}
 \caption{Radiative decays of equal mass quarkonia}
 \label{fig:1}
\end{figure}

Now, the EM transition amplitude of the process, $V\rightarrow A^+\gamma$ (where $V=1^{--}$, and $A^{++}=1^{++}$) is
\begin{equation}\label{15}
 M_{fi}=-2e\int \frac{d^4q}{(2\pi)^4}Tr[\overline{\Psi}_{A}(P', q'){\not}\epsilon^{\lambda'}\Psi_{V}(P,q)S_F^{-1}(-p_2)].
\end{equation}

where the total $M_{fi}$ is two times the contribution from Diagram 1, since both diagrams contribute equally. Here, $\overline{\Psi}_{A}(P', q')=\gamma_4\Psi_{A}^{\dag}(P', q')\gamma_4$ is adjoint 4D BS wave function for final axial meson, while $\Psi_{V}(P,q)$ is BS wave function of initial vector meson. Here we have taken, $P,q$ and $\epsilon^{\lambda}$ as the total momentum, internal momentum, and polarization vector of initial hadron, $H$, while $P',q'$, and $\epsilon^{\lambda''}$  as the corresponding variables of the final hadron, $H'$. Further, $k$, and $\epsilon^{\lambda'}$ are momentum and polarization vectors of emitted photon. Thus if $p_{1,2}$, and $p'_{1,2}$ are the momenta of the two quarks in initial and final hadron respectively, then, the momentum relations for initial hadron are: $P=p_1+p_2$, and $p_{1,2}=\frac{1}{2}P\pm q$, while for final hadron, $P'=p'_1+p'_2$, and $p'_{1,2}=\frac{1}{2}P'\pm q'$, while the relation between their internal momentum is: $q'=q-\frac{1}{2}k$ (see \cite{vaishali21} for details).

To simplify calculations, we prefer to work in the rest frame of the initial hadron, where we decompose the internal momentum $q'=(\hat{q}',iM\sigma')$ of the final hadron into two components with $ \hat{q}'=q'-\sigma'P$, and $\sigma'=\frac{q'.P}{P^2}$, with $\hat{q}'$ being transverse to the initial hadron momentum $P$, and  $\sigma'$ being longitudinal to $P$. Thus, $P.\hat{q}'=0$. The relationship between the transverse components, and longitudinal components of internal momenta of the two hadrons is given as in Eq.(13) and (15) of \cite{vaishali21}. From the Feynman diagrams we see that conservation of momentum demands that, $P= P'+k$. Now, for the first diagram, we have the kinematical relations, $p_1=p'_1+k$, and $-p_2=-p'_2$, where $k=P-P'$ is the momentum of the emitted photon. From kinematical relations, it can be shown that the energy of final meson can be expressed as \cite{bhatnagar20,vaishali21}, $E'=\frac{M^2+M'^2}{2M}$. It is to be noted that 4D BS wave functions of initial vector  meson involved in the process (in variable, $\hat{q}$) is  $\Psi_V(P,q)=S_F(p_1)\Gamma_V(\hat{q})S_F(-p_2)$, while the wave function of the final axial vector meson is, $\Psi_{A}(P',q')=S_{F}(p'_{1}) \Gamma_A(\hat{q}')S_F(-p'_{2})$, where, $\hat{q}'=q'-\frac{q'.P}{P^2}P$ is transverse to initial hadron momentum, $P$, and $\Gamma_A(\hat{q}')$ is the hadron-quark vertex function for the final meson. Now, we reduce the above equation to the effective 3D form by integrating over the longitudinal component, $Md\sigma$. This can be expressed as,

\begin{equation}\label{17}
 M_{fi}=-2e\int \frac{d^3 \hat q}{(2\pi)^3}\int\frac{iMd\sigma}{(2\pi )} Tr[\bar{\Gamma}_A(\hat q')S_F(p'_1) {\not}\epsilon'' S_F(p_1) \Gamma_V(\hat q) S_F(-p_2)].
\end{equation}

To calculate further, we split the quark and anti-quark propagators in terms of projection operators, and carrying out in integrations in complex $\sigma$ plane over the poles of the quark propagators, we obtain the expression for effective 3D form of transition amplitude, $M^{1}_{fi}$ under Covariant Instantaneous Ansatz for Diagram 1, as\cite{bhatnagar20,vaishali21},

\begin{multline}\label{31}
 M^{1}_{fi}=-ie\int \frac{d^3 \hat q}{(2\pi)^3} \frac{1}{M^2}Tr\bigg[ \alpha_1 {\not}P\overline{\psi}_A^{++}(\hat q'){\not}\epsilon''\psi_V^{++}(\hat q)
 + \alpha_2 {\not}P\overline{\psi}_A^{++}(\hat q'){\not}\epsilon''\psi_V^{--}(\hat q)\\
 +\alpha_3 {\not}P\overline{\psi}_A^{--}(\hat q'){\not}\epsilon''\psi_V^{++}(\hat q)
 + \alpha_4 {\not}P\overline{\psi}_A^{--}(\hat q'){\not}\epsilon''\psi_V^{--}(\hat q)\bigg]
 \end{multline}

Here, $\alpha_1,...,\alpha_4$ \cite{bhatnagar20,vaishali21} are the results of contour integrals over $Md\sigma$. $M_{fi}$ above is a generalized expression for transition amplitude expressed as a linear superposition of terms involving all possible combinations of $++$, and $--$ components of Salpeter wave functions of final and initial hadrons. Now on lines on \cite{wang11}, to simplify calculations, we ignore the terms involving $\psi^{+-}, \psi^{-+}$ and $\psi^{--}$ that contribute less than $1\%$, and thus consider only the terms involving $\psi^{++}$ that give dominant contribution, and write $M^{1}_{fi}$ as,

\begin{equation}
M^{1}_{fi}=-ie\int \frac{d^3 \hat q}{(2\pi)^3} \frac{1}{M^2}Tr\bigg[ \alpha_1 {\not}P\overline{\psi}_A^{++}(\hat q'){\not}\epsilon''\psi_V^{++}(\hat q)\bigg].
\end{equation}

where the $++$ components of the B.S. wave function for axial vector meson, and vector meson \cite{hluf16,wang06} respectively, can be obtained from the equations,

\begin{equation}\label{c1}
 \psi_{A}^{\pm\pm}(\hat q')= \Lambda^{\pm}_{1}(\hat q')\frac{{\not}P}{M}\psi_A(\hat q')
 \frac{{\not}P}{M}\Lambda^{\pm}_{2}(\hat q'),
\end{equation}

and

\begin{equation}\label{c1}
 \psi_V^{\pm\pm}(\hat q)= \Lambda^{\pm}_{1}(\hat q)\frac{{\not}P}{M}\psi_V(\hat q)
 \frac{{\not}P}{M}\Lambda^{\pm}_{2}(\hat q).
\end{equation}

Now, to calculate the process, we need the 3D BS wave functions, $\psi_V(\hat{q})$, and $\psi_A(\hat{q})$ for vector and axial vector mesons respectively. We then start with the general 4D form of BS wave functions \cite{smith69,alkofer02}, expressed as a superposition of various Dirac structures, each multiplied by scalar amplitudes, $f_i(P,q)$, which are all independent. We then make use of their 3D decomposition under Covariant Instantaneous Ansatz \cite{bhatnagar23,bhatnagar18,hluf16,eshete19,bhatnagar23,narang23}. We further make use of the 3D Salpeter equations that are obtained by 3D reduction of the 4D Bethe-Salpeter equation under Covariant Instantaneous Ansatz following a sequence of steps outlined in \cite{eshete19,bhatnagar23} that are expressed as,

\begin{eqnarray}
 &&\nonumber(M-\omega_1-\omega_2)\psi^{++}(\hat{q})=\Lambda_{1}^{+}(\hat{q})\Gamma(\hat{q})\Lambda_{2}^{+}(\hat{q}),\\&&
   \nonumber(M+\omega_1+\omega_2)\psi^{--}(\hat{q})=-\Lambda_{1}^{-}(\hat{q})\Gamma(\hat{q})\Lambda_{2}^{-}(\hat{q}),\\&&
\nonumber \psi^{+-}(\hat{q})=0,\\&&
 \psi^{-+}(\hat{q})=0.\label{fw5}
\end{eqnarray}

These 3D Salpeter equations in Eq.(7) depend on the Lorentz-invariant variable $\hat{q}^2=q^2-(q.P)^2/P^2$  which is a scalar, whose validity extends over the entire time-like region of the 4D space, while also keeping contact with the surface, $P.q=0$ (hadron rest frame). Putting these 3D wave functions into the last two Salpeter equations, $\Psi^{+-}(\hat{q})=0$, and  $\Psi^{-+}(\hat{q})=0$ provide constraint relations between the amplitudes of Dirac structures for any meson. Then, putting this wave function into the first two Salpeter equations leads to two coupled integral equations in two independent Salpeter amplitudes. With the structure of the interaction kernel incorporated, these coupled equations are then decoupled and reduced to two identical algebraic equations, which are in turn reduced to mass spectral equations. The 3D wave functions for $1^{--}$ and $1^{++}$ (for equal mass quarkonia) are expressed as \cite{bhatnagar23,eshete19}:

\begin{eqnarray}
&&\nonumber \psi_V(\hat q)=N_V\bigg[iM{\not}\epsilon+{\not}\epsilon {\not}P+\frac{i}{2m}{\not}P{\not}\epsilon{\not}\hat q+\frac{i}{2m}\hat q.\epsilon{\not}P\bigg]\phi_V(\hat{q});\\&&
\psi_{A}(\hat q')=N_{A}\gamma_5\bigg[iM'{\not}\epsilon'+{\not}\epsilon'{\not}P'+2i\frac{{\not}\epsilon'{\not}P'{\not}\hat q' }{M}\bigg]\phi_{A^+}(\hat q'),
\end{eqnarray}

where as mentioned above, $\epsilon^{\lambda}$ and $\epsilon^{\lambda''}$ are the polarization vectors of vector and axial vector mesons respectively. In \cite{hluf16,bhatnagar18,eshete19,vaishali21}, we had derived the analytic forms of 3D radial wave functions for $S, P, V$ and $A$ mesons from their respective mass spectral equations in an approximate harmonic oscillator basis, that were used to calculate various M1 and E1 transitions in \cite{bhatnagar20,vaishali21}. These wave functions for $V$ and $A$ quarkonia are:

\begin{equation}\label{25}
\begin{split}
   \phi_{V}(1S,\hat q)&=\frac{1}{\pi^{3/4}}\frac{1}{\beta_{V}^{3/2}}e^{-\frac{\hat q^2}{2\beta_{V}^2}}\\
 \phi_{V}(2S,\hat q)&=\sqrt{\frac{3}{2}}\frac{1}{\pi^{3/4}}\frac{1}{\beta_{V}^{3/2}}
  \bigg(1-\frac{2\hat q^2}{3\beta_{V}^2}\bigg)e^{-\frac{\hat q^2}{2\beta_{V}^2}}\\
  \phi_V(1D,\hat q)&=\sqrt{\frac{4}{15}}\frac{1}{\pi^{3/4}}\frac{1}{\beta_V^{7/2}}\hat q^2e^{-\frac{\hat q^2}{2\beta_V^2}}\\
\phi_{V}(3S,\hat q)&=\sqrt{\frac{15}{8}}\frac{1}{\pi^{3/4}}\frac{1}{\beta_{V}^{3/2}}
     \bigg(1-\frac{4\hat q^2}{3\beta_{V}^2}+\frac{4\hat q^4}{15\beta_{V}^4}\bigg)e^{-\frac{\hat q^2}{2\beta_{V}^2}}\\
   \phi_{A}(1p,\hat q)&=\sqrt{\frac{2}{3}}\frac{1}{\pi^{3/4}}\frac{1}{\beta_{A}^{5/2}}\hat q e^{-\frac{\hat q^2}{2\beta_{A}^2}}\\
 \phi_{A}(2p,\hat q)&=\sqrt{\frac{5}{3}}\frac{1}{\pi^{5/2}}\hat q (1-\frac{2\hat q^2}{5\beta_{A}^2})
e^{-\frac{\hat q^2}{2\beta_{A}^2}}\\
 \end{split}
\end{equation}
where $\beta_{A,V}$ are their inverse range parameters, with $\beta_A=(\frac{2}{3}M\omega_{q\bar{q}}^{2})^{\frac{1}{4}}$, and $\beta_V=(m\omega_{q\bar{q}}^{2})^{\frac{1}{4}}$\cite{eshete19,vaishali21a}.
\bigskip

Now, the $++$ components of vector ($1^{--}$), and axial ($1^{++}$) vector meson wave functions are:
\begin{eqnarray}
&&\nonumber \Psi_{V}^{++}(\hat{q})=N_V\phi_V(\hat{q})[i{\not}\epsilon a_1+{\not}\epsilon{\not}P a_2+i{\not}\epsilon{\not}P{\not}\hat{q}a_3+
{\not}\epsilon{\not}\hat{q}a_4+i{\not}P a_5+{\not}P{\not}\hat{q}a_6+
i{\not}\hat{q}{\not}P{\not}\epsilon a_7+i{\not}\hat{q}a_8]\\&&
\overline{\Psi}_{A}^{++}(\hat{q}')=N_A\phi_V(\hat{q}')\gamma_5[i{\not}\epsilon' b_1+{\not}\epsilon' {\not}P' b_2+i{\not}P' b_5+i{\not}P'{\not}\hat{q}'{\not}\epsilon' b_3].
\end{eqnarray}

where the coefficients associated with various Dirac structures for $V$ meson are:
\begin{eqnarray}
&&\nonumber a_1=\frac{M}{4}-\frac{2mM}{\omega}-\frac{M\hat{q}^2}{8\omega m}+\frac{m^2M}{4\omega^2}+\frac{M\hat{q}^2}{4\omega^2}-\frac{\hat{q}^2}{4\omega}\\&&
\nonumber a_2=-\frac{3m}{8\omega}+\frac{1}{4}+\frac{m^2}{4\omega^2}+\frac{\hat{q}^2}{2\omega^2}\\&&
\nonumber a_3=\frac{1}{4\omega}-\frac{\hat{q}^2}{8\omega^2m}+\frac{1}{8m}+\frac{m}{8\omega^2}\\&&
\nonumber a_4=0\\&&
\nonumber  a_5=(\epsilon.P)[\frac{\hat{q}^2}{4m\omega M}+\frac{m^2}{4\omega^2 M}-\frac{m}{2\omega M}+\frac{\hat{q}^2}{2\omega^2M}]-
(\hat{q}.\epsilon)[\frac{1}{8m}-\frac{m}{8\omega^2}-\frac{3\hat{q}^2}{8\omega^2m}]\\&&
\nonumber a_6=(\epsilon.P)[\frac{m}{4\omega^2M}+\frac{1}{2\omega M}+(\hat{q}.\epsilon)[-\frac{9}{8\omega^2}]\\&&
 \nonumber a_7=-\frac{1}{4\omega}+\frac{m}{4\omega^2}\\&&
a_8=(\hat{q}.\epsilon)[-\frac{M}{2\omega^2}+\frac{1}{4\omega}]
\end{eqnarray}

Similarly we write down the coefficients of $b_1,...,b_7$ associated with $A$ meson are:

\begin{eqnarray}
&&\nonumber b_1=\frac{M'}{4}+\frac{mM'}{2\omega'}+\frac{m^2 M'}{4\omega'^2}-\frac{M'\hat{q}'^2}{4\omega'^2}-\frac{\hat{q}'^2}{m}\\&&
\nonumber b_2=\frac{1}{4}-\frac{m^2}{4\omega'^2}+\frac{\hat{q}'^2}{4\omega'^2};~ b_3=\frac{1}{2M'}+\frac{\hat{q}'^2}{2M'\omega'^2}\\&&
\nonumber b_4=\frac{3m}{2\omega'};~ b_5=-\frac{\hat{q'}.\epsilon'}{\omega'}\\&&
 b_6=-(\hat{q}'.\epsilon')\frac{\hat{q}'^2}{M'\omega'^2};~b_7=(\hat{q}'.\epsilon')\frac{m}{M'\omega'^2}.
\end{eqnarray}

Now, our  mass spectrum, and the 3D wave functions $\phi(\hat{q})$ in Eq.(9) were calculated from the mass spectral equations, that are in turn derived from the 3D Salpeter equations in Eq.(7) in the rest frame of the hadron (please see \cite{hluf16,eshete19}.

The Bethe-Salpeter normalizers, $N_V$, and $N_A$ are obtained through current conservation condition,
\begin{equation}
2iP_{\mu}=\int \frac{d^4 q}{(2\pi)^4} Tr\bigg[\overline{\psi}(P,q)[\frac{\partial}{\partial P_{\mu}}S_F^{-1}(p_1)]\psi(P,q)S_F^{-1}(-p_2)\bigg] +(1 \leftrightarrow 2),
\end{equation}

This leads their algebraic forms,

\begin{eqnarray}
&&\nonumber N_A ^{-2} = \frac{4}{M} \int \frac{d^3 \hat{q}}{(2\pi)^3}\phi^2_A(\hat{q})[2mM-M^2+2\hat{q}^2];\\&&
N_V^{-2} = \frac{4}{M} \int \frac{d^3 \hat{q}}{(2\pi)^3}\phi^2_V(\hat{q})[-\frac{5M^2\hat{q}^2}{6m^2}+\frac{20M\hat{q}^2}{3m}].
\end{eqnarray}

Here, it is to be mentioned that, the transverse component of internal momentum of the final meson can be expressed as, $\hat q'=\hat q+ \hat{m}_2\hat{P}'$\cite{bhatnagar20}, where  $\hat{m}_2$ act as momentum partitioning parameters. Now squaring both sides of this relation, and making use of the fact that $\hat{P}'$, and $\hat{q}$ are both transverse to the initial hadron momentum, along with the momentum of emitted meson, $|\hat{P}'|=|\overrightarrow{P}|=\frac{M^2-M'^2}{2M}$, we can express the relationship between $\hat{q}'^2$, and $\hat{q}^2$ as \cite{bhatnagar20}:

\begin{equation}
\hat{q}'^2=\hat{q}^2+2\hat{m}_2 \frac{(M^2-M'^2)}{2M}|\hat{q}|+\hat{m}_2^2\frac{(M^2-M'^2)^2}{4M^2}
\end{equation}

where,$|\hat{q}|$ is the length of the 3-D vector, $\hat{q}$, defined as $|\hat{q}| =\sqrt{\hat{q}^2}=\sqrt{q^2-(q.P)^2/P^2}$, and is a Lorentz-invariant \cite{wang06, bhatnagar23}) variable.

The transition amplitude, $M_{fi}$ is expressed as,

\begin{equation}
M_{fi}=-2ie_Q\int \frac{d^3\hat{q}}{(2\pi)^3}\frac{1}{M^2}Tr[\alpha_1 {\not}P \bar{\Psi}^{++}_{A}(\hat{q}'){\not}\epsilon''\Psi^{++}_{V}(\hat{q})]
\end{equation}

The trace over the gamma matrices can be expressed as,

\begin{equation}
[Tr]=\alpha_1\bigg[X'_1 \epsilon_{\mu \nu \alpha \beta}\epsilon''_{\mu}\epsilon'_{\nu}\epsilon_{\alpha}P_{\beta}+ X'_2\epsilon_{\mu \nu \alpha \beta}k_{\mu}\epsilon_{\nu}\epsilon'_{\alpha}\epsilon''_{\beta}+ X'_3 \epsilon_{\mu \nu \alpha \beta} P_{\mu}k_{\nu}\epsilon'_{\alpha}\epsilon_{\beta}+ X'_4\epsilon_{\mu \nu \alpha \beta}k_{\mu}P_{\nu}\epsilon_{\alpha}\epsilon''_{\beta} + X'_5 \epsilon_{\mu \nu \alpha \beta} P_{\mu}k_{\nu}\epsilon'_{\alpha}\epsilon''_{\beta}\bigg],
\end{equation}

where,

\begin{eqnarray}
&&\nonumber X'_1=b_1 a_1- M^2 b_2 a_2 - (\frac{M^2}{2}-\frac{1}{2}P'.P)b_3 a_1-M^2\hat{q}^2b_3 a_3+M^2\hat{q}^2b_3 a_7 - (\frac{M^2}{2}-\frac{1}{2}P'.P)b_4a_2-\hat{q}^2b_4a_4\\&&
\nonumber X'_2=M^2b_2a_2-(M^2-\frac{1}{2}P'.P)b_3a_1+M^2\hat{q}^2b_3a_3-M^2\hat{q}^2b_3a_7+\frac{M^2}{2}b_4a_2\\&&
\nonumber X'_3=[-1 -\frac{P'.P}{M^2}]b_3a_1\\&&
\nonumber X'_4=b_6 a_1+(M^2 - \frac{P'.P}{M^2})b_7a_2-\hat{q}^2b_7a_4\\&&
X'_5=(-M^2-\frac{1}{2}P'.P)b_3a_5-\hat{q}^2b_3a_8.
\end{eqnarray}

Thus, we can write $M_{fi}$ as,

\begin{eqnarray}
&&\nonumber M_{fi}=-2ie_Q\frac{N_V N_A}{M^2}[G_1~\epsilon_{\mu \nu \alpha \beta}\epsilon''_{\mu}\epsilon'_{\nu}\epsilon_{\alpha}P_{\beta}+ G_2~\epsilon_{\mu \nu \alpha \beta}k_{\mu}\epsilon_{\nu}\epsilon'_{\alpha}\epsilon''_{\beta}+(P.\epsilon'') G_3~ \epsilon_{\mu \nu \alpha \beta} P_{\mu}k_{\nu}\epsilon'_{\alpha}\epsilon_{\beta}+\\&&
(I.\epsilon')G_4~\epsilon_{\mu \nu \alpha \beta}k_{\mu}P_{\nu}\epsilon_{\alpha}\epsilon''_{\beta} + (I.\epsilon) G_5~ \epsilon_{\mu \nu \alpha \beta} P_{\mu}k_{\nu}\epsilon'_{\alpha}\epsilon''_{\beta}],
\end{eqnarray}

where,
\begin{eqnarray}
&&\nonumber G_1=\int \frac{d^3\hat{q}}{(2\pi)^3}\alpha_1 X'_1(\hat{q})\phi_A(\hat{q})\phi_V(\hat{q})\\&&
\nonumber G_2= \int \frac{d^3\hat{q}}{(2\pi)^3}\alpha_1 X'_2(\hat{q})\phi_A(\hat{q})\phi_V(\hat{q})\\&&
\nonumber G_3=\int \frac{d^3\hat{q}}{(2\pi)^3}\alpha1 X'_3\phi_A(\hat{q})\phi_V(\hat{q})\\&&
\nonumber G_4=\int \frac{d^3\hat{q}}{(2\pi)^3}\alpha_1 X'_4\phi_A(\hat{q})\phi_V(\hat{q})\\&&
G_5=\int \frac{d^3\hat{q}}{(2\pi)^3}\alpha_1 X'_5\phi_A(\hat{q})\phi_V(\hat{q}).
\end{eqnarray}

Now, since $\hat{q}$ is an effective 3D vector, we can write, $\hat{q}.\epsilon'=|\hat{q}|(I.\epsilon')$, where we take $I_{\mu}$ as a unit vector along the direction of $\hat{q}$, and is expressed as, $I=\frac{\hat{q}}{|\hat{q}|}$, where $|\hat{q}|$ is a Lorentz-invariant variable\cite{bhatnagar23} defined above. Taking the initial vector meson to be in its rest frame, where $P=(\overrightarrow{0},iM$), and since the polarization vector of final axial meson $\epsilon''=(\overrightarrow{\epsilon}'',i0)$, due to which the dot product, $P.\epsilon''=0$. Thus the term with $G_3$ in $M_{fi}$ will not contribute. Further, $G_4$ and $G_5$ correspond to magnetic dipole $M2$ transitions, and make sub-dominant contributions to $M_{fi}$. These sub- dominant contributions are expected to contribute $10- 20\%$ \cite{ylshi16,deng16} of the decay widths, and are thus ignored in present calculation. The dominant contributions to decay widths, that correspond to $E1$ transitions, should come from $G_1$, and $G_2$ terms, in terms of which we write $M_{fi}$ as,

\begin{equation}
M_{fi}=-2ie_Q \frac{N_V N_A}{M^2}[ G_1~\epsilon_{\mu\nu\alpha\beta}~\epsilon^{\lambda''}_\mu \epsilon^{\lambda'}_\nu \epsilon^{\lambda}_\alpha P_\beta + G_2~\epsilon_{\mu\nu\alpha\beta}~\epsilon^{\lambda''}_\mu \epsilon^{\lambda'}_\nu \epsilon^{\lambda}_\alpha k_\beta].
\end{equation}

Thus, we can write $M_{fi}$ as,

\begin{eqnarray}
&&\nonumber M_{fi}=-2ie_Q \frac{N_V N_A}{M^2}\epsilon^{\lambda''}_{\mu}\epsilon^{\lambda'}_{\nu}\epsilon^{\lambda}_{\alpha}M_{\mu\nu\alpha},\\&&
M_{\mu\nu\alpha}=G_1~\epsilon_{\mu\nu\alpha\beta}P_{\beta}+G_2~\epsilon_{\mu\nu\alpha\beta}k_{\beta}.
\end{eqnarray}

Now condition for gauge invariance, $k_{\mu}M_{\mu\nu\alpha}=0$, leads to, $G_1=0$. Thus, we obtain,

\begin{equation}
M_{fi}=-2ie_Q \frac{N_V N_A}{M^2}G_2~\epsilon_{\mu\nu\alpha\beta}~\epsilon^{\lambda''}_\mu \epsilon^{\lambda'}_\nu \epsilon^{\lambda}_\alpha k_\beta.
\end{equation}

Now, to calculate the decay widths, we need to calculate the spin averaged amplitude modulus square, $|\overline{M}_{fi}|^2$, where
$|\overline{M}_{fi}|^2=\frac{1}{2j+1}\sum_{\lambda,\lambda'}|{M}_{fi}|^2$, where we average over the initial polarization states $\lambda$ of V-meson, and sum over the final polarization states, $\lambda''$ of A-meson, and $\lambda'$ of photon. We make use of the normalizations,
$\Sigma_{\lambda} \epsilon_{\mu}^{\lambda}\epsilon_{\nu}^{\lambda}=\frac{1}{3}(\delta_{\mu\nu}+\frac{P_{\mu}P_{\nu}}{M^2})$ for vector meson, $\Sigma_{\lambda''} \epsilon_{\mu}^{\lambda''}\epsilon_{\nu}^{\lambda''}=\frac{1}{3}(\delta_{\mu\nu}+\frac{P''_{\mu}P''_{\nu}}{M''^2})$ for final axial meson, and
$\Sigma_{\lambda'} \epsilon_{\mu}^{\lambda'}\epsilon_{\nu}^{\lambda'}=\delta_{\mu\nu}$, for the emitted photon, with $M_{fi}$ taken from the previous
equation.

We obtain,

\begin{equation}
|\overline{M}_{fi}|^2=\frac{4}{9}\frac{e_{Q}^{2}N_{A}^{2} N_{V}^{2}}{M^{4}}|G_2|^2\bigg[\frac{(P.k)^2}{M^2}+\frac{(P'.k)^2}{M'^2}-\frac{(P'.k)(P'.P)(P.k)}{M^2M'^2}\bigg].
\end{equation}

Taking charge of charmed quark, $e_Q=\frac{2}{3}e$, and the dot products of momenta calculated in the rest frame of initial vector meson:
$P'.P=-\frac{(M^2+M'^2)}{2}$, $P.k=-\frac{M^2-M'^2}{2}$, $P'.k=-\frac{M^2-M'^2}{2}$ we can write,

\begin{equation}
|\overline{M}_{fi}|^2=\frac{8\pi\alpha_{em}}{81} N_A^2 N_V^2|G_2|^2\frac{(M^2 - M'^2)^2(M^2 + M'^2)}{M^6 M'^2}.
\end{equation}

The decay width of the process ($V\rightarrow A\gamma$) in the rest frame of the initial vector meson is expressed as
\begin{equation}
 \Gamma_{V\rightarrow A\gamma}=\frac{|\overline{M}_{fi}|^2}{8\pi M^2}|\overrightarrow{P'}|,
\end{equation}

where we make use of the fact that modulus of the momentum of the emitted axial vector meson can be expressed in terms of masses of particles as,
$|\overrightarrow{P'}|=|\overrightarrow{k}|=\omega_k=\frac{1}{2M} (M^2-M'^2)$, where, $\omega_k$ is the kinematically allowed energy of the emitted photon.

The vector and axial vector meson masses calculated in our framework\cite{eshete19,vaishali21a} are given in Table 1.

\begin{table}[h!]
\begin{center}
\begin{tabular}{p{2cm} p{3cm}  p{3.5cm}  }
\hline\hline
 &\footnotesize{BSE-CIA}\cite{eshete19,vaishali21a}&\small Expt.\cite{zyla20} \\
\hline
\small$M_{J/\Psi}(1S)$& 3.0970& 3.0969$\pm$0.000011\\
\small$M_{\Psi}(2S)$& 3.6676& 3.6861$\pm$0.00034\\
\small$M_{\Psi}(1D)$& 3.7716& 3.7773$\pm$0.00033\\
\small$M_{\chi_{c1}}(1P)$& 3.4783&3.510$\pm$ 0.0007    \\
\small$M_{\chi_{c1}}(2P)$& 3.9560&3.871.69$\pm$ 0.00017 \\

\hline \hline
\end{tabular}
\end{center}
\caption{Mass spectra of ground and excited states of vector ($1^{--}$) and axial vector ($1^{++}$) quarkonia (in GeV) in BSE-CIA used in the transitions studied in this work, along with data}
\end{table}

The E1 decay widths for $V\rightarrow A \gamma$ involving vector ($1^{++}$), and axial vector ($1^{++}$) $c\bar c$ mesons are calculated in our model along with results of other models are given Table 2.
\begin{table}[h!]
\begin{center}
\begin{tabular}{p{4.5cm} p{3cm} p{3.1cm} p{2.4cm} p{1.8cm} }
  \hline\hline
                                                    &BSE-CIA &  Expt.\cite{workman22}  & CQM\cite{deng16}    & RQM\cite{ebert03}     \\
   \hline
 $\Gamma_{\psi(2S)\rightarrow \chi_{c1}(1P)\gamma}$&23.54    &28.665$\pm$0.78 &20 &22.9 \\
$\Gamma_{\psi(3S)\rightarrow \chi_{c1}(1P)\gamma}$&22.72    &$<$ 272$\pm$ 34 & &  \\
$\Gamma_{\psi(1D)\rightarrow \chi_{c1}(1P)\gamma}$&48.89    &67.77$\pm$2.49 &70 &  \\

 \hline\hline
 \end{tabular}
 \caption{Radiative decay widths of heavy-light $1^{--}$ charmonia (in keV) for E1 transitions in BSE, along with experimental data  and results of other models.}
 \end{center}
\end{table}

\section{Radiative transition, $A \rightarrow V\gamma$}
We now consider the radiative decay process, $A\rightarrow V\gamma$. At leading order, this process again involves a quark-triangle loop, but with two hadron-quark vertices, the initial hadron, $A=1^{++}$, and the final hadron, $V=1^{--}$. There are two diagrams that contribute to this process in Fig.2, both of which contribute equally. In each diagram, there are two Lorentz frames involved, the rest frame of the initial meson, $H$,  and the rest frame of final meson, $H'$.

As in previous section, we again write the total momentum and the internal momentum of initial hadron, $H(=A)$ as $P$, and $q$, while the corresponding momentum variables for final hadron, $H'(=V)$ as, $P'$, and $q'$. And let $k$, and $\epsilon^{\lambda'}$ be momentum and polarization vectors of emitted photon, while $\epsilon^{\lambda}$ be the polarization vector of final vector meson. Thus if $p_{1,2}$, and $p'_{1,2}$ are the momenta of the two quarks in initial and final hadron respectively, then, the kinematical relations between the momenta of initial and final hadrons can be expressed as, $q'=q-\frac{1}{2}k$ (see \cite{vaishali21} for details). To simplify calculations, we again work in rest frame of initial hadron. The amplitude for the decay process, $A\rightarrow V\gamma$, is two times the amplitude for Diagram 1. Following Section 2, we write the invariant amplitude for this process as,

\begin{equation}
 M_{fi}= -2ie_Q\int \frac{d^3 \hat q}{(2\pi)^3} \frac{1}{M^2}Tr\bigg[\alpha_1 {\not}P\overline{\psi}_V^{++} (\hat q'){\not}\epsilon''\psi_A^{++} (\hat q)\bigg],
\end{equation}

where the $++$ components of vector and axial vector meson wave functions are given as in Eq.(10) of the previous section, except that we now take $(\epsilon, P, \hat{q})$ as the polarization vector and momentum variables of initial axial meson, while, ($\epsilon'', P', \hat{q}'$) as the corresponding variables of the final vector meson.

Now trace over the Dirac gamma matrices can be expressed as,

\begin{equation}
TR=\alpha_1 [X_1 \epsilon_{\mu\nu\alpha\beta}P_{\mu}\epsilon_{\nu}\epsilon'_{\alpha}\epsilon''_{\beta}+X_2 \epsilon_{\mu\nu\alpha\beta}k_{\mu}\epsilon_{\nu}\epsilon'_{\alpha}\epsilon''_{\beta}+ X_3 \epsilon_{\mu\nu\alpha\beta}P_{\mu}k_{\nu}\epsilon''_{\alpha}\epsilon_{\beta}+ X_4 \epsilon_{\mu\nu\alpha\beta}P_{\mu}k_{\nu}\epsilon'_{\alpha}\epsilon''_{\beta} +X_5 \epsilon_{\mu\nu\alpha\beta}P_{\mu}k_{\nu}\epsilon_{\alpha}\epsilon''_{\beta}]
\end{equation}

where,

\begin{eqnarray}
&&\nonumber X_1=-a_1 b_1+M^2a_2b_2+(\hat{m}_2M^2-\frac{P'.P}{P^2}\hat{m}_2M^2)a_3b_1-M^2\hat{q}^2a_3b_3+\hat{m}_2(1-\frac{P'.P}{P^2})a_4b_2-\hat{q}^2a_4b_4+\\&&
\nonumber (\hat{m}_2M^2(1-\frac{P'.P}{P2})-\hat{q}^2M^2)a_7b_1,\\&&
\nonumber X_2=-M^2a_2b_2-\frac{P'.P}{P^2}\hat{m}_2M^2a_3b_1+M^2\hat{q}^2a_3b_3-M^2\hat{m}_2a_4b_2+M^2\hat{m}_2(-2+\frac{P'.P}{P^2})a_7b_1+M^2\hat{q}^2a_7b_3,\\&&
\nonumber X_3=2\hat{m}_2[2-\frac{P'.P}{P^2}]a_3b_1,\\&&
\nonumber X_4=\hat{q}^2a_3b_5+M^2\hat{m}_2\frac{P'.P}{P^2}a_3b_6-\hat{q}^2a_7b_5-M^2\hat{m}_2(1-\frac{P'.P}{P^2})a_7b_6,\\&&
X_5=-2M^2\hat{m}_2a_6b_2+\hat{q}^2a_6b_4+\hat{m}_2a_8b_1,
\end{eqnarray}

where the coefficients, $a_1,...,a_8$, and $b_1,...,b_7$ for $A\rightarrow V\gamma$ are of the same form as in Eqs.(11)-(12) for $V\rightarrow A\gamma$ in previous section, except that, we have to interchange $M\rightleftharpoons M', \omega \rightleftharpoons \omega', \hat{q}\rightleftharpoons \hat{q}'$.

Thus, we can write $M_{fi}$ as,

\begin{eqnarray}
&&\nonumber M_{fi}=-2ie_Q\frac{N_A N_V}{M^2}[F_1 \epsilon_{\mu\nu\alpha\beta}P_{\mu}\epsilon_{\nu}\epsilon'_{\alpha}\epsilon''_{\beta}+F_2 \epsilon_{\mu\nu\alpha\beta}k_{\mu}\epsilon_{\nu}\epsilon'_{\alpha}\epsilon''_{\beta}+(P.\epsilon') F_3 \epsilon_{\mu\nu\alpha\beta}P_{\mu}k_{\nu}\epsilon''_{\alpha}\epsilon_{\beta}+(I.\epsilon)F_4\epsilon_{\mu\nu\alpha\beta}P_{\mu}k_{\nu}\epsilon'_{\alpha}\epsilon''_{\beta}+\\&& \nonumber (I.\epsilon')F_5 \epsilon_{\mu\nu\alpha\beta}P_{\mu}k_{\nu}\epsilon_{\alpha}\epsilon''_{\beta}];\\&&
\nonumber F_1=\int \frac{d^3\hat{q}}{(2\pi)^3}\alpha_1 X_1(\hat{q})\phi_A(\hat{q})\phi_V(\hat{q})\\&&
\nonumber F_2= \int \frac{d^3\hat{q}}{(2\pi)^3}\alpha_1 X_2(\hat{q})\phi_A(\hat{q})\phi_V(\hat{q})\\&&
\nonumber F_3=\int \frac{d^3\hat{q}}{(2\pi)^3}\alpha_1 X_3\phi_A(\hat{q})\phi_V(\hat{q})\\&&
\nonumber F_4=\int \frac{d^3\hat{q}}{(2\pi)^3}\alpha_1 X_4\phi_A(\hat{q})\phi_V(\hat{q})\\&&
F_5=\int \frac{d^3\hat{q}}{(2\pi)^3}\alpha_1 X_5\phi_A(\hat{q})\phi_V(\hat{q}).
\end{eqnarray}

As mentioned earlier, since $\hat{q}$ is an effective 3D variable, we have written $\hat{q}.\epsilon'=\hat{q}_{\nu}\epsilon'_{\nu}=|\hat{q}|(I.\epsilon')$, where, $I_{\mu}$ is a unit vector along the direction of $\hat{q}_{\nu}$, and is expressed as, $I=\frac{\hat{q}}{|\hat{q}|}$, where $|\hat{q}|$ is a Lorentz-invariant variable \cite{bhatnagar20,wang06,bhatnagar23}. Similarly we write $\hat{q}.\epsilon''=|\hat{q}|(I.\epsilon'')$, and $\hat{q}.\epsilon=|\hat{q}|(I.\epsilon)$.

In the above expression for $M_{fi}$ the term associated with the form factor, $F_3$ vanishes on account of $P.\epsilon'=0$, since $P=(\overrightarrow{0},iM)$ is momentum of initial axial vector meson at rest, while $\epsilon'=(\overrightarrow{\epsilon'},i0)$ is the polarization vector of the emitted photon, that is transversely polarized. Further, we wish to mention that the amplitudes, $F_4$ and $F_5$ contribute to M2 transitions, and make sub-dominant contributions to decay width, and are ignored. And the dominant contributions arise from E1 transitions that come from the amplitudes, $F_1$ and $F_2$, in terms of which we can write the amplitude, $M_{fi}$ as,

\begin{equation}
M_{fi}=-2ie_Q\frac{N_A N_V}{M^2}[F_1 \epsilon_{\mu\nu\alpha\beta}\epsilon''_\mu\epsilon'_\nu\epsilon_\alpha P_\beta + F_2 \epsilon_{\mu\nu\alpha\beta}\epsilon''_\mu\epsilon'_\nu\epsilon_\alpha k_\beta]
\end{equation}

Thus, we can write, $M_{fi}$ as,

\begin{eqnarray}
&&\nonumber M_{fi}=\epsilon''_\mu\epsilon'_\nu\epsilon_\alpha M_{\mu\nu\alpha};\\&&
M_{\mu\nu\alpha}=F_1\epsilon_{\mu\nu\alpha\beta}P_{\beta}+F_2\epsilon_{\mu\nu\alpha\beta}k_{\beta}.
\end{eqnarray}

Now, electromagnetic gauge invariance demands,
\begin{equation}
k_{\mu}M_{\mu\nu\alpha}=0,
\end{equation}

which leads to the form factor, $F_1=0$. Thus we can express the amplitude, $M_{fi}$, that corresponds to E1 transition and makes dominant contribution as,

\begin{eqnarray}
&&\nonumber M_{fi}=-2ie_Q\frac{N_A N_V}{M^2}[F_2 \epsilon_{\mu\nu\alpha\beta}\epsilon''_\mu\epsilon'_\nu\epsilon_\alpha k_\beta]\\&&
F_2= \int \frac{d^3\hat{q}}{(2\pi)^3}\alpha_1 X_2(\hat{q})\phi_A(\hat{q})\phi_V(\hat{q}).
\end{eqnarray}

Here, $X_2$ appearing in the expression for $F_2$ is given in Eqs.(21). Now, to calculate the decay widths, we need to calculate the spin averaged amplitude modulus square, $|\overline{M}_{fi}|^2$, where
$|\overline{M}_{fi}|^2=\frac{1}{2j+1}\sum_{\lambda,\lambda'}|{M}_{fi}|^2$, where we average over the initial polarization states $\lambda$ of A-meson, and sum over
the final polarization $\lambda'$ of photon. We make use of the normalizations,
$\Sigma_{\lambda} \epsilon_{\mu}^{\lambda}\epsilon_{\nu}^{\lambda}=\frac{1}{3}(\delta_{\mu\nu}+\frac{P_{\mu}P_{\nu}}{M^2})$ for vector meson, and
$\Sigma_{\lambda'} \epsilon_{\mu}^{\lambda'}\epsilon_{\nu}^{\lambda'}=\delta_{\mu\nu}$, for the emitted photon, with $M_{fi}$ taken from the previous
equation, we get, $\sum_{\lambda'}\sum_{\lambda}|\epsilon^{\lambda'}.\epsilon^{\lambda}|^2 =1$.

We obtain,

\begin{equation}
|\overline{M}_{fi}|^2=\frac{4e_Q^2N_A^2 N_V^2}{9M^{4}}|F_2|^2\bigg[\frac{(P'.k)^2}{M'^2}+\frac{(P.k)^2}{M^2}-\frac{(P'.k)(P'.P)(P.k)}{M^2M'^2}\bigg],
\end{equation}

where, the dot products of momenta calculated in the rest frame of initial vector meson are:
$P'.P=-\frac{(M^2+M'^2)}{2}$, $P.k=-\frac{M^2-M'^2}{2}$, $P'.k=-\frac{M^2-M'^2}{2}$. Thus we notice that the general expression for $|\bar{M}_{fi}|^2|$ for the process $1^{++}\rightarrow 1^{--}\gamma$ is the same as the corresponding expression for the process $1^{--}\rightarrow 1^{++}\gamma$ in Eq.(23), and (24), except for the difference in the form factors $F_2$, and $G_2$ respectively.

The decay width of the process ($A^+\rightarrow V\gamma$) in the rest frame of the initial meson is expressed as in Eq.(25), and we make use of the fact that modulus of the momentum of the emitted vector meson can be expressed in terms of masses of particles as,
$|\overrightarrow{P'}|=|\overrightarrow{k}|=\omega_k=\frac{1}{2M} (M^2-M'^2)$ (where $M/M'$ is mass of initial axial/final vector meson, where, $\omega_k$ is the kinematically allowed energy of the emitted photon.

\bigskip

\begin{table}[h!]
\begin{center}
\begin{tabular}{p{4.5cm} p{3cm} p{3.1cm} p{2.4cm} p{1.8cm} p{1.6cm} }
  \hline\hline
                                                    &BSE-CIA &  Expt.\cite{workman22} & QM\cite{deng16}    & LFM\cite{ylshi16}  &BSE\cite{wang11}   \\
   \hline
 $\Gamma_{\chi_{c1}(1P)\rightarrow \gamma J/\psi(1S)}$&291.11&288$\pm$11.52 &275 &324$\pm$20 &306 \\
$\Gamma_{\chi_{c1}(2P)\rightarrow \gamma \psi(2S)}$&69.703& 53.55$\pm$23.8   &    &    & 146.0\\
$\Gamma_{\chi_{c1}(2P)\rightarrow \gamma J\psi(1S)}$&18.959& 9.52$\pm$0.84   &    &    &33.0 \\

 \hline\hline
 \end{tabular}
 \caption{Radiative decay widths of $1^{++}$ charmonia (in keV) in present model along with experimental data and results of other models.}
 \end{center}
\end{table}

\section{Conclusion}
In present work, we have studied the radiative E1 transitions $^3P_1\rightarrow ^3S_1\gamma$, and $^3S_1\rightarrow ^3P_1\gamma$ in the framework of Bethe-Salpeter equation and calculated their decay widths. The 3D Salpeter wave functions $\phi_V(\hat{q})$, and $\phi_A(\hat{q})$ used for calculation of the transition amplitudes, were derived as analytic solutions of mass spectral equations for vector and axial quarkonia\cite{eshete19,vaishali21a} in an approximate harmonic oscillator basis. Our results of decay widths are compared with data and other models. In our framework, ratio of decay widths, $\frac{\Gamma_{\chi_{c1}(2P)\rightarrow \gamma \psi(2S)}}{\Gamma_{\chi_{c1}(2P)\rightarrow \gamma J\psi(1S)}}=3.676$ (c.f. Expt.= 5.625\cite{workman22}). The corresponding ratio of these decay widths in \cite{wang11} is 4.424.

{50}


\begin{thebibliography}{50}

\bibitem{workman22} R. L. Workman et al. (Particle Data Group), PTEP2022, 083C01 (2022).

\bibitem{zyla20} P.A.Zyla et al., (Particle Data Group), Prog. Theo. Expt. Phys. 2020, 083C01 (2020).

\bibitem{olive14} K.A.Olive \textit{et al.}, (Particle Data Group), Chin. Phys. C{\bf38}, 090001 (2014).

\bibitem{bhardwaj13} V. Bhardwaj et al. (Belle), Phys. Rev. Lett. 111, 032001 {2013}.

\bibitem{coan06} T.E.Coan et al.,  hep-ex/0509030.
\bibitem{ablikim15} M. Ablikim et al. (BESIII), Phys. Rev. Lett. 115, 011803(2015); arXiv:1503.08203 [hep-ex].

\bibitem{ablikim18} M.Ablikim et al., Phys. Rev. D97, 032008 (2018).

\bibitem{vaishali21a} V.Guleria, S.Bhatnagar, Intl. J. Theor. Phys. 60, 3143 (2021).

\bibitem{vaishali21} V.Guleria, E.Gebrehana, S.Bhatnagar, Phys. Rev. D104, 094045 (2021).
\bibitem{ylshi16} Y.L.Shi, arxiv:1611.03712[hep-ph].
\bibitem{deng16} Wei-Jun Deng, Li-Ye Xiao, Long-Cheng Gui, and Xian-Hui Zhong, arxiv: 1510.08269[hep-ph].

\bibitem{wang11} T.H. Wang, G.L. Wang, Phys. Lett. B697, 233 (2011).

\bibitem{ebert03} D. Ebert, R. N. Faustov and V. O. Galkin, Phys. Rev. D67, 014027 (2003).
\bibitem{godfrey15} S. Godfrey and K. Moats, Phys. Rev. D 92, 054034(2015).
\bibitem{segovia16}  J. Segovia, P. G. Ortega, D. R. Entem and F. Fernandez, Phys. Rev. D 93, 074027 (2016).
\bibitem{he21} J.K. He, C. J. Fan, Phys. Rev. D103, 114006 (2021).
\bibitem{he19} J.K. He, Y. D. Yang, Nucl. Phys. B943, 114627 (2019).

\bibitem{bhatnagar20} S.Bhatnagar, E.Gebrehana, Phys. Rev. D102, 094024 (2020).


\bibitem{smith69} C. H. L. Smith, Ann. Phys. (N.Y.) 53, 521 (1969).
\bibitem{alkofer02} R. Alkofer, L.V.Smekel, Phys. Rep. 353, 281 (2002).
\bibitem{bhatnagar18} S.Bhatnagar, L. Alemu, Phys. Rev. D97, 034021 (2018).
\bibitem{bhatnagar23} S.Bhatnagar, V.Guleria, Nucl. Phys. A1041, 122783 (2024).
\bibitem{narang23} M.Narang, S.Bhatnagar, Few-Body Syst. (2023).




\bibitem{hluf16} H.Negash, S.Bhatnagar, Intl. J. Mod. Phys. E25, 1650059 (2016).
\bibitem{eshete19} E.Gebrehana, S.Bhatnagar, H.Negash, Phys. Rev. D100, 054034 (2019).

\bibitem{wang06} C.H.Chang, J.K.Chen, G.L.Wang, Commun. Theo. Phys.(Beijing) 46, 467 (2006).

\bibitem{ylshi17} Y.L.Shi, Eur. Phys. J. C77, 253 (2017).

























\end{thebibliography}
\end{document}